\documentclass[sigconf]{acmart}
\copyrightyear{2023}
\acmYear{2023}
\setcopyright{acmlicensed}
\acmConference[CIKM '23] {Proceedings of the 32nd ACM International Conference on Information and Knowledge Management}{October 21--25, 2023}{Birmingham, United Kingdom.}

\acmBooktitle{Proceedings of the 32nd ACM International Conference on Information and Knowledge Management (CIKM '23), October 21--25, 2023, Birmingham, United Kingdom}
\acmPrice{15.00}
\acmISBN{979-8-4007-0124-5/23/10}
\acmDOI{10.1145/XXXXXX.XXXXXX}
\usepackage{multirow}
\usepackage{caption}
\usepackage{subcaption}
\usepackage{graphicx}
\usepackage{float} 
\usepackage{enumitem}
\usepackage{amsmath}

\AtBeginDocument{%
  \providecommand\BibTeX{{%
    \normalfont B\kern-0.5em{\scshape i\kern-0.25em b}\kern-0.8em\TeX}}}

\acmSubmissionID{1457}

\begin{document}

%%
%% The "title" command has an optional parameter,
%% allowing the author to define a "short title" to be used in page headers.
\title{ Enhancing Repeat-Aware Recommendation from a Temporal-Sequential Perspective}

%%
%% The "author" command and its associated commands are used to define
%% the authors and their affiliations.
%% Of note is the shared affiliation of the first two authors, and the
%% "authornote" and "authornotemark" commands
%% used to denote shared contribution to the research.
\author{Shigang Quan}
% \authornote{Both authors contributed equally to this research.}

% \orcid{1234-5678-9012}
% \author{G.K.M. Tobin}
% \authornotemark[1]
\affiliation{%
  \institution{Shanghai Jiao Tong University}
  \city{Shanghai}
  \country{China}
}
\email{quan123@sjtu.edu.cn}

\author{Shui Liu}
\affiliation{%
  \institution{Alibaba Group}
  \city{Hangzhou}
  \state{Zhejiang}
  \country{China}
}
\email{shui.lius@alibaba-inc.com}

\author{Zhenzhe	Zheng}
\affiliation{%
  \institution{Shanghai Jiao Tong University}
  \city{Shanghai}
  \country{China}
}
\email{zhengzhenzhe@sjtu.edu.cn}

\author{Fan	Wu}
\affiliation{%
  \institution{Shanghai Jiao Tong University}
  \city{Shanghai}
  \country{China}
}
\email{fwu@cs.sjtu.edu.cn}

%%
%% By default, the full list of authors will be used in the page
%% headers. Often, this list is too long, and will overlap
%% other information printed in the page headers. This command allows
%% the author to define a more concise list
%% of authors' names for this purpose.
\renewcommand{\shortauthors}{Trovato and Tobin, et al.}

%%
%% The abstract is a short summary of the work to be presented in the
%% article.
\begin{abstract}
Repeat consumption, such as repurchasing items and relistening songs, is a common scenario in daily life. To model repeat consumption, the repeat-aware recommendation has been proposed to predict which item will be re-interacted based on the user-item interactions. In this paper, we investigate various inherent characteristics to enhance the repeat-aware recommendation. Specifically, we explore these characteristics from two aspects: one is from the temporal aspect where we consider the time interval relationship in the user behavior sequence; the other is from the sequential aspect where we consider the sequential-level relationship in the user behavior sequence. And our intuition is that both the \textit{temporal pattern} and \textit{sequential pattern} will reflect users' intentions of repeat consumption.
 
  By utilizing these two patterns, a novel model called \textbf{T}emporal and \textbf{S}equential repeat-aware \textbf{Rec}ommendation(\textbf{TSRec} for short) is proposed to enhance repeat-aware recommendation. TSRec has three main components: 1) User-specific Temporal Representation Module (UTRM), which encodes and extracts user historical repeat temporal information. 2)Item-specific Temporal Representation Module (ITRM), which incorporates item time interval information as side information to alleviate the data sparsity problem of user repeat behavior sequence. 3) Sequential Repeat-Aware Module (SRAM), which represents the similarity between the user's current and the last repeat sequences. Extensive experimental results on three public benchmarks demonstrate the superiority of TSRec over state-of-the-art methods. The implementation code is available \footnote{\url{https://anonymous.4open.science/r/TSRec-2306/}}.

% To address the first challenge, we first define two new patterns related to repeat consumption - \textit{periodic pattern} and \textit{sequential pattern}. In SPRec, a Sequential Repeat-Aware Module (SPAM) aims to extract the \textit{sequential pattern} between the current repeat sequence and the last repeat sequence, while the target attention layer aims to model the \textit{periodic pattern} in the user's historical behavior sequence. To address the second challenge, we define the Repeat Period Matrix (RPM) from the item perspective to alleviate the data sparsity from the user perspective. Extensive experiments on public datasets demonstrate the superiority of our model and proposed sub-modules, compared with the state-of-the-art benchmarks.

\end{abstract}

%%
%% The code below is generated by the tool at http://dl.acm.org/ccs.cfm.
%% Please copy and paste the code instead of the example below.
%%


%%
%% Keywords. The author(s) should pick words that accurately describe
%% the work being presented. Separate the keywords with commas.

%% A "teaser" image appears between the author and affiliation
%% information and the body of the document, and typically spans the
%% page.
% \begin{teaserfigure}
%   \includegraphics[width=\textwidth]{sampleteaser}
%   \caption{Seattle Mariners at Spring Training, 2010.}
%   \Description{Enjoying the baseball game from the third-base
%   seats. Ichiro Suzuki preparing to bat.}
%   \label{fig:teaser}
% \end{teaserfigure}

% \received{20 February 2007}
% \received[revised]{12 March 2009}
% \received[accepted]{5 June 2009}

%%
%% This command processes the author and affiliation and title
%% information and builds the first part of the formatted document.
\maketitle

\section{Introduction}
\begin{figure*}[]
  \centering
  \includegraphics[width=0.9\linewidth]{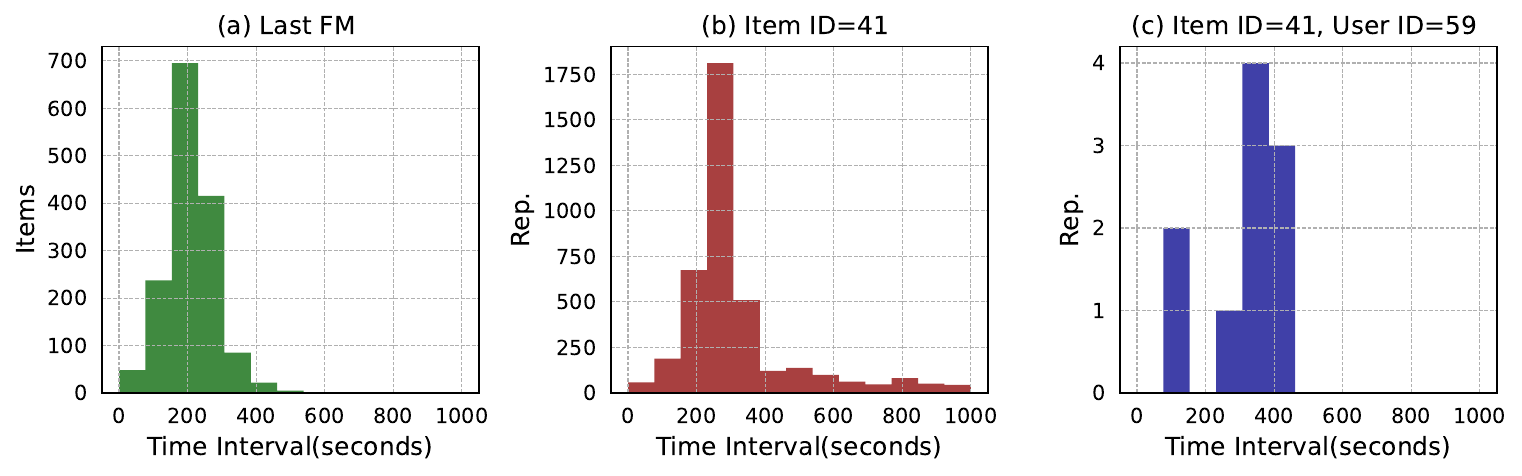}
  \caption{Data analyses to illustrate the inherent characteristics related to repeat consumption from the temporal aspect on the LastFM dataset. The x-axis is the repeat time interval of an item and the y-axis is the repeat frequency for different time intervals. }
  \label{fig: frequency}
\end{figure*}
The recommendation system has achieved great success due to its ability to capture dynamic user interests across various online platforms, such as E-commerce \cite{FPMC}, social media \cite{socialmedia2015,socialmedia}, music recommendation \cite{musicrec, ACT-R}, etc. The fundamental objective of a recommendation system is to predict the user's next item by estimating the similarity between the target item and the user's previously interacted items.

 Recently, rather than just modeling user behavior, the repeat-aware recommendation task has been proposed to explore repeat consumption, a phenomenon where users repeatedly purchase or click on certain items. Modeling users' repeat behavior can improve the accuracy of next-item prediction. To this end, some researchers attempt to model users' temporal behavior using distribution models, such as the Negative Binomial Distribution model \cite{buyit2018} and Point Process Model \cite{ SLRC}. However, these methods only focus on modeling user repeat behavior and neglect user dynamic preferences for new items, thus limiting the applicability and generalizability to scenarios with relatively low repeat ratios. Conversely, other methods, as mentioned in \cite{RepeatNet, recency, ReCANet}, aim to improve upon these limitations by incorporating an additive neural component to predict the intention of user repeat consumption, or by introducing a supervised loss, with a ground truth repeat label, to regulate model training. These methods are characterized by the absence of  modeling intentions behind repeat consumption, which impedes their effectiveness in striking a balance between recommending new and repeat items.
 
In this paper, we investigate various inherent characteristics to enhance the repeat-aware recommendation. Specifically, we explore these characteristics from two aspects. One is from the temporal aspect: we consider the time interval relationship in the user behavior sequence; the other is from the sequential aspect: we consider the sequential-level relationship in the user behavior sequence. And our intuition from these two aspects is as follows: 
% we attempt to model the behavior of user repeat consumption from the temporal-sequential perspective and to decouple the underlying intentions behind user repeat consumption. To demonstrate the temporal and sequential characteristics of user repeat consumption, we first conducted a preliminary experiment on the LastFM dataset  as illustrated in Figure \ref{fig: frequency}. And we list motivations of enhancing repeat-aware recommendation from the following two perspectives:
\begin{itemize}[leftmargin=*]
    \item \textbf{Intuition 1:} for repeatedly consumed products, users typically exhibit a fixed consumption cycle for these products. For example, after users consume milk, they are likely to repurchase it during a certain period, based on their practical demand. 
    \item \textbf{Intuition 2:} we may infer the likelihood of users making repeat consumption based on the inherent correlation within a subset of user behavior sequences. For example, when people listen to music, they tend to repeat a series of songs rather than a single track.
\end{itemize}

The above intuitions motivate us to conduct preliminary data analysis on public datasets. After preliminary data analysis, we identified patterns in the data that align with the intuitions mentioned above from two aspects: for intuition 1, we calculated the distribution of item repeat frequency with fixed repeat time intervals on the LastFM dataset in Figure \ref{fig: frequency}(a) and the proportion of these items is 13.15\%. The \textit{temporal pattern} is that repeat consumption typically exhibits repeat behavior within a defined time interval; for intuition 2, we calculated the Jaccard similarity of item sets between the user behavior sequence preceding the item and the last repeat behavior sequence of the target item across three datasets and found a relatively high degree of item overlap (45\%-60\%), as shown in Table \ref{tbl:jacc}. The \textit{sequential pattern} is that repeat consumption typically manifests as a high correlation across the entire sequence.

\begin{table}[]
 \caption{The calculated Jaccard similarity on three different datasets. It is derived by calculating the union and intersection between the user's current interacted item sequence sets and the former repeat interacted item sequence sets. }
    \centering
    \begin{tabular}{cccc}
    \toprule
    {\textbf{Dataset}}&{\textbf{RetailRocket}}&{\textbf{LastFM}}&{\textbf{Diginetica}}\\
    \midrule
    {\textbf{Jacc.(\%)}}&{49.69}&{58.10}&{45.23}\\ 
    \bottomrule
    \end{tabular}
    \label{tbl:jacc}
\end{table}

In this paper, we propose a novel model called \textbf{T}emporal and \textbf{S}equential repeat-aware \textbf{Rec}ommendation (\textbf{TSRec} for short) to automatically model repeat consumption from temporal and sequential aspects. The variants of TSRec include User-specific Temporal Representation Module (UTRM), Item-specific Temporal Representation Module (ITRM), and Sequential Repeat-Aware Module (SRAM). Specifically, to capture the \textit{temporal pattern}, we design a UTRM based on the target attention layer to encode and extract user historical repeat temporal information. But the above module suffers from data sparsity issues, because changes in the user's preferences may lead to a low frequency of repeat consumption of specific items. As depicted in Figure \ref{fig: frequency}(b), even users exhibiting high-frequency repeat consumption behavior (e.g., User ID=59) have a relatively low overall frequency of repeat consumption for a target item (Item ID=41), with a frequency of nearly 10. Therefore we propose UTRM, which incorporates item time interval information from other users as side information. Our preliminary experiment in Figure \ref{fig: frequency} has shown that the item repeat time intervals for different users were relatively uniform. Based on this finding, we define the Repeat Time Interval Matrix based on the time intervals of the target item and design an external ITRM to encode the item-specific temporal information. To represent the user multi-scale temporal information, we refer to the idea of "Network In Network" (NIN) \cite{nin} and design ITRM, comprising CNN and pooling layers, to extract local and global information. To capture the \textit{sequential pattern}, we design the SRAM, which consists of the Sequence Encoder and Sequence Matching Module, to represent the similarity between the user's current sequence and the former sequence. The Sequence Encoder will encode these two sequences through a hierarchical Bi-GRU \cite{GRU} and one-dimension convolution\cite{1DCNN} layer. Then following the "matching-aggregation" \cite{esim, esim1} framework widely used in natural language processing (NLP) tasks, we feed the encoded sentence representations into the Sequence Matching Module, which consists of several complex alignment operations, to enhance the model's ability to extract and fuse mutual relation between two sentences. Finally, we evaluate the effectiveness of the proposed TSRec on three public datasets with different characteristics. To provide benchmarks for comparison, we consider several well-known sequential and repeat-aware recommendation methods. The empirical results show that TSRec can significantly outperform all the benchmarks in terms of all the evaluation metrics.

Our major contributions can be highlighted as follows:
\begin{itemize}[leftmargin=*]
    \item A novel model TSRec is proposed to enhance repeat-aware recommendation. TSRec is the first to incorporate inherent patterns of repeat consumption  from temporal and sequential aspects into the design of the repeat-aware recommendation. 
    \item To extract the \textit{temporal pattern} and \textit{sequential pattern}, we have devised two innovative sub-modules: UTRM and SRAM, respectively
    Additionally, we have designed an ITRM sub-module to address the data sparsity issue in one user's repeat consumption sequence.
    \item We conducted extensive experiments on three publicly available datasets to validate the superiority of our proposed method over various state-of-the-art recommendation methods. Furthermore, ablation and case studies provide additional evidence of the benefits of our model.
\end{itemize}

\begin{table}[]
 \caption{Some notations and descriptions used in the paper. }
    \centering
    \resizebox{0.98\linewidth}{!}{
    \begin{tabular}{ll}
    \toprule
    {\textbf{Notation}}&{\textbf{Description}}\\
    \midrule
    {$\mathcal{U}=\{ u \}$}&{the whole user set }\\ 
    {$\mathcal{I}=\{ i \}$}&{the whole item set }\\ 
    {$\mathcal{S}^{u}=\{ i^u \}$}&{the interacted item sequence of user $u$}\\
    {$\mathcal{T}^{u}=\{ t^u \}$}&{the timestamp sequence of user $u$ corresponding $\mathcal{S}^{u}$}\\ 
    {$\mathcal{P}^{u,i}=\{ p^{u,i} \}$}&{the time interval sequence of user $u$ for item $i$}\\ 
    \midrule
    {$d \in \mathbb{N}$}&{the dimensionality of the latent representations}\\
    {$L \in \mathbb{N}$}&{the length of truncated sequence}\\
    {$\boldsymbol{e}^{u} \in \mathbb{R}^{d}$}&{the embedded user vector }\\    
    {$\boldsymbol{e}^{v} \in \mathbb{R}^{d}$}&{the embedded target item vector }\\      
    {$\boldsymbol{e}^{t} \in \mathbb{R}^{d}$}&{the embedded target time interval vector}\\ 
    {$\mathbf{M}^i\in \mathbb{N}^{m \times n}$}&{the Repeat Time Interval Matrix of item $i$}\\
    {$\mathbf{E}^{h} \in \mathbb{R}^{L \times d}$}&{the embedded historical time interval matrix}\\ 
    {$\mathbf{E}^b \in \mathbb{R}^{L \times d}$}&{the embedded user behavior matrix}\\   
    {$\mathbf{E}^l \in \mathbb{R}^{L \times d}$}&{the embedded former repeat behavior matrix}\\     
    \bottomrule
    \end{tabular}
    }
    \label{tbl: notation}
\end{table}
\section{Preliminaries}
If we denote the user and item sets as $\mathcal{U}$ and $\mathcal{I}$, respectively, their corresponding sizes can be represented by $|\mathcal{U}|$ and $|\mathcal{I}|$. In sequential recommendation, the interaction sequence of user $u \in \mathcal{U}$ is organized chronologically as $\mathcal{S}^{u}=\{{i^{u}_1},i^{u}_2,...,i^{u}_{|\mathcal{S}|}\}$ and time sequence is $\mathcal{T}^{u}=\{{t^{u}_1},t^{u}_2,...,t^{u}_{|\mathcal{S}|}\}$, which means the user $u$ interacts with the item $i^u$ at time $t^u$.\\
\textbf{Task of repeat-aware recommendation.} Given $\mathcal{U}, \mathcal{I}, \mathcal{S}$ and $\mathcal{T}$, the task of repeat-aware recommendation is to predict the score that user $u$ clicks on the next (either new or repeat) item  $i$ at time $t^{u}_{|\mathcal{S}|+1}$:\\
\begin{equation}
    \hat{y}^{u}_{t_{|\mathcal{S}|+1}}=p(i^u_{|\mathcal{S}|+1}|\mathcal{S}^{u},\mathcal{T}^u)
\end{equation}
\begin{definition}
\textbf{Repeat Consumption.} Given $\mathcal{U}, \mathcal{I}, \mathcal{S}$ and $\mathcal{T}$, repeat consumption exists in user $u$ if and only if user $u$'s interacted item $i^u_{|\mathcal{S}|+1} = i^u \in \mathcal{S}^{u}$.
\end{definition}

\section{Methodology}
\begin{figure*}[h]
  \centering
  \includegraphics[width=\linewidth]{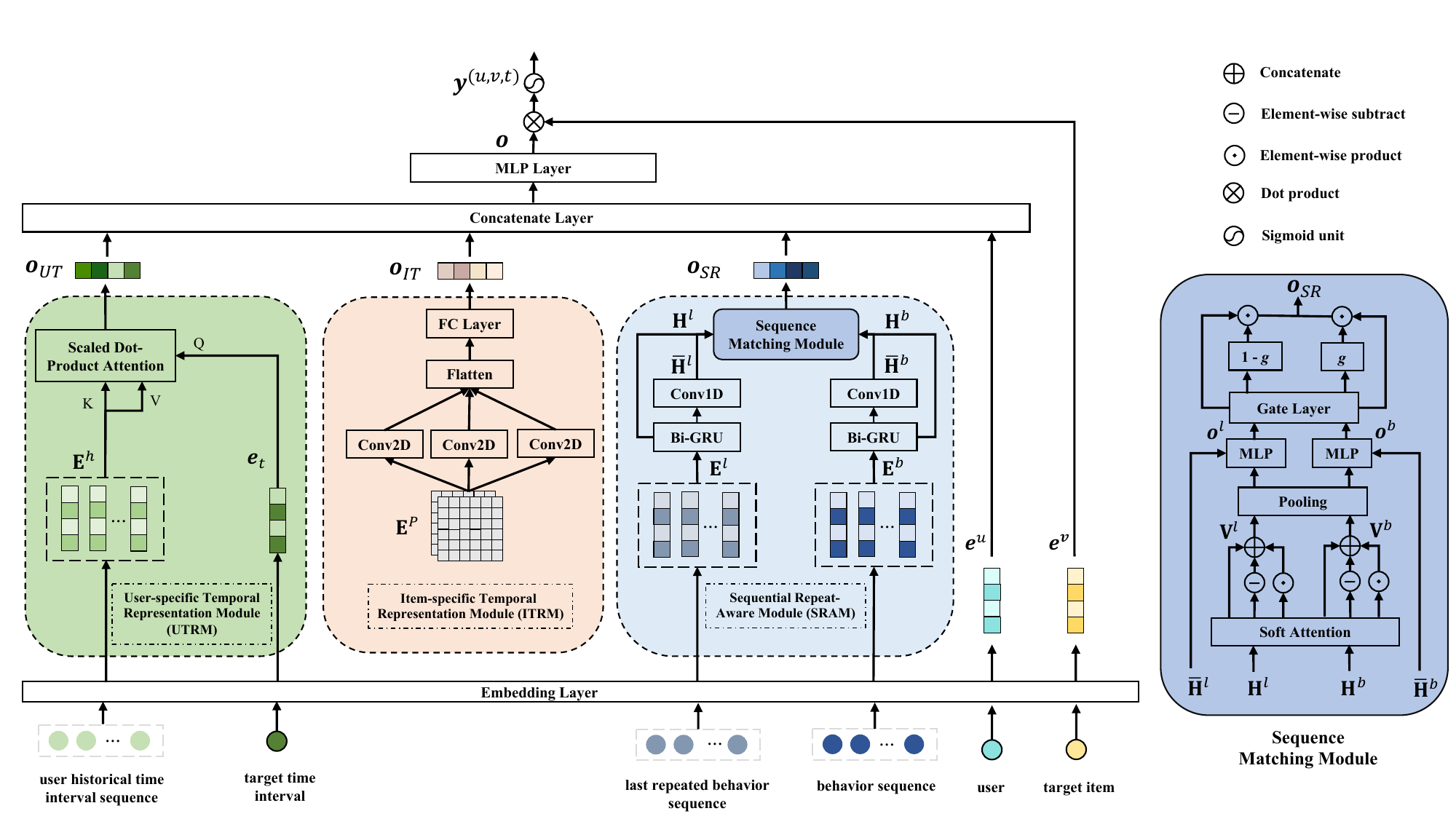}
  \caption{The overview architecture of TSRec, which consists of three parts: User-specific Temporal Representation Module (UTRM), Item-specific Temporal Representation Module (ITRM), and Sequential Repeat-Aware Module (SRAM). The UTRM captures \textit{temporal pattern} by modeling the relation of repeat time intervals in user sequence. The ITRM captures the item-specific \textit{temporal pattern} by extracting the temporal information in Repeat Time Interval Matrix. The SRAM first encodes the user's current behavior sequence and former behavior sequence and then aligns these two encoded vectors. All these representations will be integrated into the prediction layer for the next-item prediction.}
  \label{fig: model}
\end{figure*}

In this section, we propose the TSRec in Figure \ref{fig: model} to model repeat consumption in the sequential recommendation. TSRec consists of three sub-modules: User-specific Temporal Representation Module (UTRM), Item-specific Temporal Representation Module (ITRM), and Sequential Repeat-Aware Module (SRAM). And representation vectors from these three sub-modules will be concatenated with embedded user vectors as the user's complete preference for the target item.
We list some key notations and their descriptions in Table \ref{tbl: notation}.

\subsection{Feature Representation}
\subsubsection{Sequence Embedding}
We first transform the $u$'s interacted item sequence $\mathcal{S}^{u}=\{{i^{u}_1},i^{u}_2,\dots, i^{u}_{|\mathcal{S}|}\}$ into a fixed-length sequence $\boldsymbol{s}=\{{i^{u}_{|\mathcal{S}|-L+1}},i^{u}_{|\mathcal{S}|-L+2},\dots,$ $i^{u}_{|\mathcal{S}|}\}$, where $L$ is the predefined sequence length. If the length of $\mathcal{S}^{u}$ exceeds $L$, we keep only the most recent $L$ behaviors. Conversely, if the sequence length is less than $L$, we insert padding items to the left until the desired length $L$ is reached. For each item $ i \in \mathcal{I}$, we first transform id categories into sparse one-hot vectors. Then we transform these high-dimensional representations into dense vectors as the embedded vectors.

For $\mathcal{S}^{u}=\{{i^{u}_1},i^{u}_2,\dots,i^{u}_{j-1},i^{u}_{j},\dots,i^{u}_{k-1},i^{u}_{k},\dots,i^{u}_{|\mathcal{S}|}\}$, if $i^{u}_{j}=i^{u}_{k}$, we define $\mathcal{S}^{u,b}=\{i^{u}_{j+1},i^{u}_{j+2},\dots,i^{u}_{k-1}\}$ as the user behavior sequence preceding the target item $i^{u}_{k}$, and $\mathcal{S}^{u,l}=\{i^{u}_{1},i^{u}_{2},\dots,i^{u}_{j-1}\}$ as the last repeat behavior sequence of the target item $i^{u}_{k}$. As previous operation, $\mathcal{S}^{u,b}, \mathcal{S}^{u,l}$ will be cut off and padded into fixed length $L$ as $\boldsymbol{v}^b, \boldsymbol{v}^l$. And these two sequence vectors will be transformed into two embedded matrices by utilizing the embedding layer, named $\mathbf{E}^b,\mathbf{E}^l \in \mathbb{R}^{L \times d}$.
\subsubsection{Time Interval Embedding}
To represent the temporal information of repeat consumption, we need to pre-process the timestamp sequence $\mathcal{T}^{u}$ and transform it into the learnable matrix. We first define the Repeat Time Interval Matrix as follows:
\vspace{-0.5em}
\begin{definition}
\label{RPM}
\textbf{Repeat Time Interval Matrix.} For user $u$, the timestamp sequence of the interacted item $i$  can be denoted as $\mathcal{T}^{u,i}=\{t^{u,i}_1,t^{u,i}_2,..., t^{u,i}_n, t^{u,i}_{n+1}\}$. The $k$-th time interval of item $i$ is $p^{u,i}_{k}=\Big\lfloor\frac{|t^{u,i}_{k+1}-t^{u,i}_{k}|}{p_{\min }}\Big\rfloor$($k=1,2,\cdots,n$), where $p_{\min} $ is the minimal time interval and $p^{u,i}_{k}$ is a discrete value. The time interval sequence of user $u$ for item $i$ is $\mathcal{P}^{u,i} = \{p^{u,i}_{1}, p^{u,i}_{2},...,p^{u,i}_{n}\}$. The Repeat Time Interval Matrix of item $i$ can be defined as $\mathbf{M}^i \in \mathbb{N}^{m \times n}$:
\begin{equation}
    \mathbf{M}^i=\left[\begin{array}{c}
\mathcal{P}^{u_1,i} \\[0.2cm]
\mathcal{P}^{u_2,i} \\[0.2cm]
\cdots \\[0.2cm]
\mathcal{P}^{u_m,i}
\end{array}\right]=
\left[\begin{array}{cccc}
p_{1}^{u_1,i} & p_{2}^{u_1,i} & \ldots & p_{n}^{u_1,i} \\[0.1cm]
p_{1}^{u_2,i} & p_{2}^{u_2,i} & \ldots & p_{n}^{u_2,i} \\[0.1cm]
\vdots & \vdots & \ddots & \vdots  \\[0.1cm]
p_{1}^{u_m,i} & p_{2}^{u_m,i} & \ldots & p_{n}^{u_m,i}
\end{array}\right].
\end{equation}
\end{definition}
Like sequence embedding, we can convert discrete time interval values into one-hot vectors and then transform them into dense embedded vectors by the embedding layer. Specifically, the target time interval (time interval between current timestamp and last behavior's timestamp) $p^{u,i}$, user historical time interval matrix $\mathcal{P}^{u, i}$, and Repeat Time Interval Matrix $\mathbf{M}^i$ will be embedded by using the embedding layer named $\boldsymbol{e}^t \in \mathbb{R}^{ d}$, $\mathbf{E}^{h} \in \mathbb{R}^{n \times d}$ and  $\mathbf{E}^{P} \in \mathbb{R}^{m \times n \times d}$.
\subsection{User-specific Temporal Representation Module}
The proposed module aims to extract the \textit{temporal pattern} from user historical repeat consumption behavior primarily through the target attention layer. To achieve this, we first obtain the target time interval embedding vector $\boldsymbol{e}^t$ and historical time interval embedding matrix $\mathbf{E}^h=\{\boldsymbol{e}^h_1,\cdots, \boldsymbol{e}^h_k,\cdots, \boldsymbol{e}^h_n\}$ through the embedding layer. We can figure out if the target item is the clicked item of the user, the target time interval $\boldsymbol{e}^t$ will be the next element of $\mathbf{E}^h$. We explicitly incorporate temporal information into the attention procedure, utilizing the widely used attention methodology - \textit{Scaled Dot-Product Attention} \cite{Transformer}. The \textit{Scaled Dot-Product Attention} computes weights of the sequence representations of user historical time intervals, after converting them by three learnable weight matrices, $\mathbf{W}^Q,\mathbf{W}^K,\mathbf{W}^V \in \mathbb{R}^{d \times d}$, as:
\begin{align}
    \operatorname{sim}(\boldsymbol{e}^t, \boldsymbol{e}^h_{k})&=\frac{(\mathbf{W}^Q\boldsymbol{e}^t) \cdot(\mathbf{W}^K \boldsymbol{e}^h_{k})^\mathsf{T}}{\sqrt{d}},\\
    \operatorname{Attention}(\boldsymbol{e}^t, \boldsymbol{e}^h_{k})&=\frac{\exp (\operatorname{sim}(\boldsymbol{e}^t, \boldsymbol{e}^h_{k}))}{\sum_{k^{\prime}=1}^n \exp (\operatorname{sim}(\boldsymbol{e}^t, \boldsymbol{e}^h_{k^{\prime}}))},\\
    \boldsymbol{o}_{UT}=\sum_{k=1}^n &\operatorname{Attention}(\boldsymbol{e}^t, \boldsymbol{e}^h_{k}) \cdot(\mathbf{W}^V \boldsymbol{e}^h_{k}),
\end{align}
where $\boldsymbol{o}_{UT} \in \mathbb{R}^d$ denotes the weighted sum pooling of the historical time interval embedding matrix. By performing the aforementioned operations, we can capture the \textit{temporal pattern} in the user's historical sequence and their personalized repeat consumption intention for different items.

\subsection{Item-specific Temporal Representation Module}
The user historical repeat consumption sequence for item $i$ is inherently sparse, necessitating the incorporation of other user time intervals as side information. Therefore, we define the Repeat Time Interval Matrix for target item $i$ in definition \ref{RPM}. However, capturing the complex temporal repeat consumption information in user behavior is challenging. To address this, we draw inspiration from the "Network In Network" (NIN) \cite{nin, ResNeSt} concept and propose a multi-scale feature extraction approach. We design $C=c_1,c_2,\dots,c_k,\dots, c_n$ two-dimensional convolutional filters with different kernel sizes to extract \textit{temporal patterns} of different scales and aggregate these representations using both mean-pooling and max-pooling operations:
\begin{align}
    \boldsymbol{m}^{c_k}_{\text {mean}}&= \operatorname{Mean\_Pooling}(\operatorname{Conv2D}^{c_k}(\mathbf{E}^{P})),\\
    \boldsymbol{m}^{c_k}_{\text {max}}&= \operatorname{Max\_Pooling}(\operatorname{Conv2D}^{c_k}(\mathbf{E}^{P})).
\end{align}
The pooling representation vectors of the $c$-th two-dimensional convolutional filter $\boldsymbol{m}^{c_k} =[\boldsymbol{m}^{c_k}_{\text {mean}}; \boldsymbol{m}^{c_k}_{\text {max}}]$. We aggregate these representation vectors of different scales as follows:
\begin{align}
    \boldsymbol{o}_{IT} &=\operatorname{MLP}([\boldsymbol{m}^{c_1};\dots;\boldsymbol{m}^{c_k};\dots;\boldsymbol{m}^{c_n}]),
\end{align}
where $\boldsymbol{o}_{IT} \in \mathbb{R}^d$.
\subsection{Sequential Repeat-Aware Module}
\subsubsection{Sequence Encoder}
After the embedding layer, we obtain the user behavior matrix $\mathbf{E}^b=\{\boldsymbol{e}^b_1,\boldsymbol{e}^b_2,\dots,\boldsymbol{e}^b_L\}$ and the last repeat behavior matrix $\mathbf{E}^l=\{\boldsymbol{e}^l_1,\boldsymbol{e}^l_2,\dots,\boldsymbol{e}^l_L\}$, which exhibit strong temporal relations in Table \ref{tbl:jacc}. To encode these two embedded matrices, a simple pooling operation is intuitive but this encoding operation may lose the local temporal information and can not reflect user dynamic preferences. In this paper, we adopt the stacked encoded layers: Bidirectional Gated Recurrent Unit (Bi-GRU) \cite{GRU} and one-dimension convolution \cite{1DCNN}, which extract different features from multi-dimensional perspectives. Bi-GRU focuses on sequence and context features, while also being able to memorize long-term dependencies through its use of gating mechanisms. Also, Bi-GRU can processes input data from both forward and backward directions simultaneously, enabling better understanding and representation of the input data. In contrast, one-dimensional convolution is better at extracting local patterns across the entire sequence, regardless of position, through its use of sliding windows. The formulations of the forward GRU encoder for $\mathbf{E}^l=\{\boldsymbol{e}^l_1,\boldsymbol{e}^l_2,\dots,\boldsymbol{e}^l_j,\dots, \boldsymbol{e}^l_L\}$ are listed as follows:
\begin{align}
 \boldsymbol{r}_j &= \sigma(\mathbf{W}_r \boldsymbol{e}^{l}_j+\mathbf{U}_r \overrightarrow{\boldsymbol{h}}^{l}_{j-1}), \\
 \boldsymbol{z}_j &= \sigma(\mathbf{W}_z \boldsymbol{e}^{l}_j+\mathbf{U}_z \overrightarrow{\boldsymbol{h}}^{l}_{j-1}), \\
 \tilde{\boldsymbol{h}}_j  &= \tanh (\mathbf{W}_h \boldsymbol{e}^l_j+\mathbf{U}_h (\boldsymbol{r}_j \odot \tilde{\boldsymbol{h}}_{j-1})), \\
 \overrightarrow{\boldsymbol{h}}^{l}_j &= \boldsymbol{z}_j \odot \overrightarrow{\boldsymbol{h}}^{l}_{j-1}+(1-\boldsymbol{z}_j) \odot \tilde{\boldsymbol{h}}_j,
\end{align}
where $\sigma$ denotes the sigmoid function, $\mathbf{W}_{r}, \mathbf{W}_{z}, \mathbf{W}_{h},\mathbf{U}_r,\mathbf{U}_z, \mathbf{U}_h \in \mathbb{R}^{d \times d}$ are learnable weight matrices, $\boldsymbol{r}_j$ is the reset gate, $\boldsymbol{z}_j$ is the update gate and $\tilde{\boldsymbol{h}}_j$ is the candidate activation. The $j$-th hidden state of the Bi-GRU layer is obtained by concatenating the hidden state in the forward GRU and the backward GRU as $\boldsymbol{h}^{l}_j =[\overrightarrow{\boldsymbol{h}}^{l}_j; \overleftarrow{\boldsymbol{h}}^{l}_j]$ and the encoded representation matrix is $\mathbf{H}^l=\{\boldsymbol{h}^{l}_1, \boldsymbol{h}^{l}_2, \dots, \boldsymbol{h}^{l}_j, \dots, \boldsymbol{h}^l_L\}$. Like the encoding operation of the Bi-GRU layer for $\mathbf{E}^l$,  we can obtain the Bi-GRU encoded representation matrix for $\mathbf{E}^l$, named $\mathbf{H}^b$. To extract local information of $\mathbf{H}^{l}, \mathbf{H}^{b} \in \mathbb{R}^{d \times L}$, we use one-dimensional convolutional encoders, and these convolutional encoders will use the sliding windows of different lengths on $\mathbf{H}^b, \mathbf{H}^l$:
\begin{align}
    \bar{\mathbf{H}}^b &= \operatorname{Conv1D}(\mathbf{H}^b)\\
    \bar{\mathbf{H}}^l &= \operatorname{Conv1D}(\mathbf{H}^l)
\end{align}
\subsubsection{Sequence Matching Module}
The Bi-GRU encoded representation matrices $\mathbf{H}^l, \mathbf{H}^b \in \mathbb{R}^{d \times L}$ need to be aligned and extracted the relation between the above two parts. In this paper, we adopt the soft attention alignment \cite{soft_attention_1,soft_attention_2}, with the alignment matrix computed as follows:
\begin{equation}
    \mathbf{A}={(\mathbf{H}^l)^{\mathsf{T}}}\cdot{\mathbf{H}}^b,
\end{equation}
where each element in $\mathbf{A}=\{a_{jk}\} \in \mathbb{R}^{L \times L}$ represents the alignment score of the element between $\mathbf{H}^l$ and $\mathbf{H}^b$. To get the alignment representation, the encoded matrix will be transformed by the weighted sum of the alignment matrix:
\begin{align}
 \tilde{\boldsymbol{h}}^b_{k} &=\sum_{k=1}^{L} \frac{\exp (a_{j k})}{\sum_{j^{\prime}=1}^{L} \exp (a_{j^{\prime} k})}\boldsymbol{h}^l_{j},\\
 \tilde{\boldsymbol{h}}^l_{j} &=\sum_{j=1}^{L} \frac{\exp (a_{j k})}{\sum_{k^{\prime}=1}^{L} \exp (a_{j k^{\prime}})}\boldsymbol{h}^b_{k},
\end{align}
where the aligned matrices are $\tilde{\mathbf{H}}^l, \tilde{\mathbf{H}}^b \in \mathbb{R}^{d \times L}$. Intuitively, $\tilde{\mathbf{H}}^b$ is a weighted sum of $\mathbf{H}^l$ with the weights being determined by the similarity score between $\mathbf{H}^b$ and $\mathbf{H}^l$. To further enhance the alignment information, we employ the element-wise subtract and product operation between $\tilde{\mathbf{H}}^b$ and $\mathbf{H}^b$, as well as $\tilde{\mathbf{H}}^l$ and $\mathbf{H}^l$. We expect that such operations could help sharpen local information between elements in the sentence representations and capture relationships such as contradiction. The representations of element-wise subtract and product operations will be concatenated with encoded and aligned representations as follows:
\begin{align}
\mathbf{V}^{b} &=  \operatorname{MLP}([\mathbf{H}^b; \tilde{\mathbf{H}}^{b}; \mathbf{H}^b - \tilde{\mathbf{H}}^{b}; \mathbf{H}^b\odot \tilde{\mathbf{H}}^{b}]), \\
\mathbf{V}^{l} &=  \operatorname{MLP}([\mathbf{H}^l; \tilde{\mathbf{H}}^{l}; \mathbf{H}^l - \tilde{\mathbf{H}}^{l}; \mathbf{H}^l\odot \tilde{\mathbf{H}}^{l}]), 
\end{align}
where $\mathbf{V}^{l}\in \mathbb{R}^{d \times L}, \mathbf{V}^{b} \in \mathbb{R}^{d \times L}$, and $\operatorname{MLP}$ is a Multi-Layer Perception \cite{mlp}. The above operations aim to transform the representations from different perspectives. These matrix representations will be projected into vectors by mean-pooling and max-pooling operations:
\begin{align}
\label{equ:9}
\boldsymbol{v}^{b}_{\text {mean}}=\sum_{j=1}^{L} \frac{\mathbf{V}^{b}}{L}&,  \boldsymbol{v}^{b}_{\text {max}}=\max _{j=1}^{L} \mathbf{V}^{b},  \\
\label{equ:10}
\boldsymbol{v}^{l}_{\text {mean}}=\sum_{j=1}^{L} \frac{\mathbf{V}^{l}}{L}&,  \boldsymbol{v}^{l}_{\text {max}}=\max _{j=1}^{L} \mathbf{V}^{l},
\end{align}
where $\boldsymbol{v}^{l}_{\text {mean}}, \boldsymbol{v}^{l}_{\text {max}}, \boldsymbol{v}^{b}_{\text {mean}}, \boldsymbol{v}^{b}_{\text {max}}  \in \mathbb{R}^{d}$. We flatten the encoded matrices $\bar{\mathbf{H}}^l$, $\bar{\mathbf{H}}^b$ as $\bar{\boldsymbol{h}}^l , \bar{\boldsymbol{h}}^b \in \mathbb{R}^{d \times L}$ and concatenate them with $\boldsymbol{v}^{l}_{\text {mean}}$ $, \boldsymbol{v}^{l}_{\text {max}}, \boldsymbol{v}^{b}_{\text {mean}}, \boldsymbol{v}^{b}_{\text {max}}$ as two vectors:
\begin{align}
\boldsymbol{o}^b&=\operatorname{MLP}( [\boldsymbol{v}^{b}_{\text {mean}} ; \bar{\boldsymbol{h}}^b; \boldsymbol{v}^{b}_{\text {max}}]), \\
\boldsymbol{o}^l&=\operatorname{MLP}( [\boldsymbol{v}^{l}_{\text {mean}} ; \bar{\boldsymbol{h}}^l; \boldsymbol{v}^{l}_{\text {max}}), 
\end{align}
To fuse $\boldsymbol{o}^l$ and $\boldsymbol{o}^b$, we refer to the gated-based fusion operation \cite{fusion} and obtain the fused vectors:
\begin{align}
    \boldsymbol{g} &= \sigma(\mathbf{W}_1 \boldsymbol{o}^{l}+ \mathbf{W}_2 \boldsymbol{o}^{b}), \\
       \boldsymbol{o}_{SR} &= \boldsymbol{o}^{l} \odot \boldsymbol{g} + \boldsymbol{o}^{b} \odot ( \boldsymbol{1}-\boldsymbol{g} ), 
\end{align}
where $\mathbf{W}_1\in \mathbb{R}^{d \times d}, \mathbf{W}_2 \in \mathbb{R}^{d \times d}$ are learnable weight matrices of the gate layer and $\sigma$ is the sigmoid activation function. $\boldsymbol{g} \in \mathbb{R}^{d}$ is a vector rather than a scalar, which enables the neural gate to regulate the contributions of $\boldsymbol{o}^l$ and $\boldsymbol{o}^b$ at the dimension granularity.

\subsection{Network Training}
After different modules, we get different representation vectors $\boldsymbol{o}_{UT}, \boldsymbol{o}_{IT}, \boldsymbol{o}_{SR}$. These vectors, together with user embedding vector $\boldsymbol{e}^{u}$, will be combined as:
\begin{equation}
 \boldsymbol{o}=\operatorname{MLP}([\boldsymbol{o}_{UT}; \boldsymbol{o}_{IT};\boldsymbol{o}_{SR}; \boldsymbol{e}^{u}]),
\end{equation}
where $\boldsymbol{o} \in \mathbb{R}^{d}$ denotes the fused representation. Finally, we adopt a dot-product of $\boldsymbol{o}$ with the target item $\boldsymbol{e}^v$ as a normalized prediction score of user $u$ rating for target item $v$ at the timestamp $t$:
\begin{equation}
    \boldsymbol{y}^{(u,v,t)} = \sigma(\boldsymbol{o}^{\mathsf{T}}\cdot \boldsymbol{e}^v)).
\end{equation}
In this paper, since the length of user sequences may vary, we adopt a sliding window strategy \cite{CASER}. This approach involves dividing each sequence into windows of size $L$ if the sequence length exceeds $L$, or padding the sequence if the length is less than $L$. If $\Tilde{\mathcal{T}}^{u}=\{L+1, L+2,...,|\mathcal{S}_t^u|\}$ is the timestamp set we need to predict, our optimized objective loss for this prediction task can be defined with Cross-Entropy loss as follows:
\begin{equation}
        \mathcal{L} = \sum_u \sum_{t\in \tilde{\mathcal{T}}^u }- \log(\boldsymbol{y}^{(u, v^{+},t)})+\sum_u \sum_{t\in \Tilde{\mathcal{T}}^u }\sum_{v^{-}} -(1-\log(\boldsymbol{y}^{(u, v^{-},t)}).
\end{equation}
Here, $v^{+}$ represents the positive sample where user $u$ interacted at timestamp $t$, and $v^{-}$ represents the randomly sampled negative sample that does not exist in the positive sample set.

\section{EXPERIMENTS}
\begin{table*}[]
 \caption{ Performance comparison of all baseline methods. The best results of all methods are indicated in bold, while the second best results are indicated in underlined. Improvements over baselines are statistically significant with p < 0.01.}
    \centering

    \renewcommand{\arraystretch}{1.2}
    \resizebox{0.98\linewidth}{!}{
    \begin{tabular}{l|l|cccccc|cccc|cc}
    % \toprule
    \hline  \textbf{ Datasets } & \textbf{ Metric } & \textbf{ PopRec } & \textbf{ GRU4Rec } & \textbf{ NARM } & \textbf{ Caser } & \textbf{ SASRec } & \textbf{ BERT4Rec }  & \textbf{ RepeatNet } & \textbf{ SLRC$\_$NCF} & \textbf{ ReCANet } & \textbf{ NovelNet } & \textbf{TSRec} & \textbf{Improv.} \\
    \hline  \multirow{6}{*}{\textbf{ RetailRocket }} & \textbf{ HR@5 } & 0.3265 & 0.6923 & 0.7276 & 0.6545 & \underline{0.7798} & 0.7729 & 0.5418 & 0.6412 & 0.5149 & {0.6934} & \textbf{0.8112}  & {$4.03\%$}  \\
     {} &\textbf{ HR@10 } & 0.4667 & 0.7662 & 0.7980 & 0.7368 & \underline{0.8307} & 0.8244 & 0.6408 & 0.7002 & 0.6097 & {0.7679} & \textbf{0.8620}& $3.77\%$  \\
     {} &\textbf{ HR@20 } & 0.6369 & 0.8415 & 0.8669 & 0.8198 & \underline{0.8849} & 0.8808 & 0.7575& 0.7679 & 0.7329 & {0.8435} & \textbf{0.9087} & $2.69\%$ \\
    {} &\textbf{ NDCG@5 } & 0.2249 & 0.6057 & 0.6388 & 0.5525 & \underline{0.7125} & 0.7103 & 0.4656 & 0.5867 & 0.4466 & {0.5977} & \textbf{0.7240} & {$0.91\%$}   \\
     {} &\textbf{ NDCG@10 } & 0.2701 & 0.6296 & 0.6616 & 0.5791 & \underline{0.7310} & 0.7270 & 0.4975 & 0.6058 & 0.4772 & {0.6218} & \textbf{0.7405} & {$0.89\%$}  \\
     {} &\textbf{ NDCG@20 } & 0.3130 & 0.6486 & 0.6790 & 0.6001 & \underline{0.7427} & 0.7412 & 0.5269 & 0.6229 & 0.5082& {0.6410} & \textbf{0.7523} & {$0.62\%$}   \\
    \hline  \multirow{6}{*}{\textbf{ LastFM}} & \textbf{ HR@5 } & 0.4452 & 0.8162 & 0.8298 & 0.8460 & \underline{0.8543} & 0.8484 & 0.7407 & 0.8541 & 0.8460 & 0.8110 & \textbf{0.8660} &{$1.37\%$}  \\
     {} &\textbf{ HR@10 } & 0.5766 & 0.8816 & 0.8891 & 0.8982 & \underline{0.9045} & 0.8940 & 0.8140 & 0.8993 & 0.8934 & 0.8750 & \textbf{0.9128} & {$0.92\%$}  \\
     {} &\textbf{ HR@20 } & 0.7051 & 0.9292 & 0.9334 & 0.9237 & \underline{0.9418} & 0.9328 & 0.8791 & 0.9392 & 0.9269 & 0.9246 & \textbf{0.9508} & {$1.22\%$}\\
    {} &\textbf{ NDCG@5 } & 0.3216 & 0.7223 & 0.7451 & 0.7397 & 0.7754 & 0.7775& 0.6675 & 0.7776 & \underline{0.7808} & 0.7246 & \textbf{0.7903} & {$1.54\%$}  \\
     {} &\textbf{ NDCG@10 } & 0.3641 & 0.7435 & 0.7644 & 0.7568& 0.7918& 0.7923 & 0.6913 & 0.7923& \underline{0.7935} & 0.7554 & \textbf{0.8056} & {$1.52\%$} \\
     {} &\textbf{ NDCG@20 } & 0.3966 & 0.7557 & 0.7757 & 0.7655 & 0.8013 & 0.8021 & 0.7078 & 0.8025&\underline{0.8026}& 0.7580  & \textbf{0.8152} & {$1.57\%$} \\
     \hline  \multirow{6}{*}{\textbf{ Diginetica}} & \textbf{ HR@5 } & 0.2142 & 0.7460 & \underline{0.8899} & 0.8315 & 0.8810 & 0.8410 & 0.6392 & 0.6890 & 0.2703& {0.8840} & \textbf{0.9386} &{$5.47\%$}  \\
     {} &\textbf{ HR@10 } & 0.3268 & 0.8348 & \underline{0.9300} & 0.8837 & 0.9165 & 0.8907 & 0.7519 & 0.7462& 0.3743 & {0.9269} & \textbf{0.9575} & {$2.96\%$}  \\
     {} &\textbf{ HR@20 } & 0.4764 & 0.9004 & \underline{0.9573} & 0.9243 & 0.9435 & 0.9284 & 0.8123 & 0.8055& 0.5151 & {0.9536} & \textbf{0.9699} & {$1.32\%$}\\
    {} &\textbf{ NDCG@5 } & 0.1418 & 0.6070& 0.7609 & 0.7391 & \underline{0.7848} & 0.7373 & 0.4068 & 0.6172 & 0.2044 & {0.7471} & \textbf{0.8054} & {$2.62\%$}  \\
     {} &\textbf{ NDCG@10 } & 0.1780 & 0.6359 & 0.7740& 0.7561 & \underline{0.7964}& 0.7535 & 0.4441 & 0.6358 & 0.2378 & {0.7612} & \textbf{0.8116} & {$1.91\%$} \\
     {} &\textbf{ NDCG@20 } & 0.2156& 0.6525 & 0.7810 & 0.7664 & \underline{0.8032} & 0.7630 & 0.4594 & 0.6508 &0.2732 & {0.7679} & \textbf{0.8147} & {$1.43\%$} \\
    \bottomrule
    \end{tabular}
  }
    \label{tbl: baseline}
\end{table*}
In this section, our objective is to address the following research questions (RQs):
 \begin{itemize}[leftmargin=*]
     \item \textbf{RQ1}: How does TSRec perform compared to various state-of-the-art recommendation methods with different settings?
     \item \textbf{RQ2}: How effective are the key modules (e.g., User-specific Temporal Representation Module, Item-specific Temporal Representation Module) in TSRec? 
     \item \textbf{RQ3}: How do different hyper-parameters, such as embedding layer dimension $d$ and user sequence length $L$ , affect the TSRec's performance?
     \item \textbf{RQ4}: How is the capture capability of TSRec considering the temporal information modeling?
 \end{itemize}

\subsection{Experimental Settings}
\begin{table}[]
 \caption{ Statistics of our experimented datasets. Inter. means the number of interactions; Avg. means the average number of interactions per user;  Rep.(\%) calculates the ratio of repeated user-item interactions. }
    \centering
    \begin{tabular}{cccccc}
    \toprule
    {\textbf{Dataset}}&{\textbf{Users}}&{\textbf{Items}}&{\textbf{Inter.}}&{\textbf{Avg.}}&{\textbf{Rep.(\%)}}\\
    \midrule
    {\textbf{RetailRocket}}&{217,508}&{51,586}&{910,947}&{4.00}&{35.03}\\ 
    {\textbf{LastFM}}&{728}&{11,543}&{1,108,999}&{15.41}&{89.29}\\ 
    {\textbf{Diginetica}}&{187,938}&{24,700}&{861,859}&{4.59}&{16.53}\\ 
    \bottomrule
    \end{tabular}
    \label{tbl: dataset}
\end{table}
\subsubsection{\textbf{Datasets.}} We conduct our experiments on the following datasets:
\begin{itemize}[leftmargin=*]
    \item \textbf{RetailRocket\footnote{\url{https://www.kaggle.com/retailrocket/ecommerce-dataset}}} This dataset is generated from an online shopping site-Retailrocket.  It is a collection of user browsing activities on an online shopping site spanning six months.
    \item \textbf{LastFM\footnote{\url{https://www.lastfm.com}}} The LastFM dataset is a rich collection of music listening histories generated by users of the LastFM platform. We employ the 1K version in our experiments.
    \item \textbf{Diginetcia\footnote{\url{http://cikm2016.cs.iupui.edu/cikm-cup}}} The Diginetica dataset, which was released by CIKM Cup 2016, is a collection of users' clicked stream data captured in an e-commerce website. We only used the transaction data that was released.
\end{itemize}

 Following previous work \cite{SASREC, BERT4REC, GRU4REC}, we only consider implicit feedback (click/order) and we treat explicit feedback (ratings) as implicit feedback in the datasets. We group the interaction records for each user and build the interaction sequence for each user by sorting these interaction records according to the timestamps. For each dataset, we divide the data by time into a training set (70\%), validation set (10\%), and testing set (20\%). To address the cold-start issue, we filter out items that have been interacted with fewer than 10 times and only retain users whose behavior sequences consist of more than one interaction. The detailed statistics of experimental datasets are shown in Table \ref{tbl: dataset}. 
\subsubsection{\textbf{Evaluation Metrics.}} Following \cite{SASREC, BERT4REC}, we use the \textit{Top-K Hit Rate} ({HR@K}) and \textit{Top-K Normalized Discounted Cumulative Gain} (NDCG@K), which are commonly used in related research \cite{SASREC,BERT4REC}. Our reported results are based on {HR@}\{5, 10, 20\} and {NDCG@}\{5, 10, 20\}. We calculate all metrics based on item ranking and report the average scores. we conduct a paired t-test, with statistical significance being indicated when the p-value is below 0.01.
\subsubsection{\textbf{Baselines.}}
To evaluate the superiority and effectiveness of TSRec, we benchmark it with two types of baselines:\\
\textbf{General Sequential Recommendation Methods.}
\begin{itemize}[leftmargin=*]
    \item {\textbf{PopRec}}. It recommends items to users based on their ranked frequency interacted by users.
    \item {\textbf{GRU4Rec}} \cite{GRU4REC}. This method utilizes the gated recurrent unit as a sequence encoder to learn user dynamic preference with a ranking-based loss.
    \item {\textbf{NARM}} \cite{NARM}. It  integrates the RNN layers with the attention mechanism to capture the local and global preferences of users.
    \item {\textbf{Caser}} \cite{CASER}. This sequential recommendation method integrates the convolutional neural layers from both vertical and horizontal views to encode user time-varied behavior sequences.
    \item {\textbf{SASRec}} \cite{SASREC}. This method utilizes self-attention mechanisms to automatically learn long-term dependencies in the sequence. It also uses techniques such as negative sampling and masked loss to improve the training efficiency and recommendation performance of the model.
    \item {\textbf{BERT4Rec}} \cite{BERT4REC}. It is a recommendation model based on the Bidirectional Encoder Representations from Transformers (BERT) architecture. The model is optimized with the Cloze objective and has produced state-of-the-art performance among sequential learning-based baselines.
\end{itemize}
\textbf{Repeat-aware Recommendation Methods.}
\begin{itemize}[leftmargin=*]
    \item {\textbf{RepeatNet}} \cite{RepeatNet}. RepeatNet is an attention-based model, which aims to address the problem of repeat consumption. In addition to the standard attention mechanism used in sequence-to-sequence models, RepeatNet also employs a repeat-aware attention mechanism to model the probability of a user revisiting an item in the sequence.
    \item {\textbf{SLRC$\_$NCF}} \cite{SLRC}. SLRC\_NCF is proposed with a combination of two techniques: Collaborative Filtering \cite{cf} and Hawkes Process \cite{hawkes}. The model adaptively incorporates two item-specific temporal dynamics of repeat consumption into the kernel function of the Hawkes Process. 
    \item {\textbf{ReCANet}} \cite{ReCANet}. ReCANet is a novel neural architecture designed for Next Basket Recommendation (NBR) in grocery shopping, with a specific focus on repeat items.
    \item {\textbf{NovelNet}} \cite{NovelNet}. NovelNet is a recommendation model designed specifically for online novels, which encodes user interactions by taking into account fine-grained interaction attributes. It utilizes a pointer network with pointwise and listwise loss functions to model user repeat consumption behavior.
\end{itemize}
\subsubsection{\textbf{Hyper-parameter Settings.}}
Hyper-parameters are usually established by default baselines. For sequential recommendation methods, we adopt the preceding research of \cite{SASREC} and apply Adam with a learning rate of 0.001 to balance gradients. The batch size is 256, and the dropout rate is set at 0.5. We utilize the activation functions of sigmoid and ReLU. The number of two-dimensional convolutional layers $C=3$, with kernel sizes of $1 \times 1$, $1 \times 3$, and $3 \times 1$. The grid search method is applied to find the optimal settings of hyper-parameters using the validation set. These include the latent representation dimension $d$ from $\{10, 20, 50, 100, 200\}$, and the sequence length $L$ is from $\{5, 10, 15, 20\}$. To construct the Repeat Time Interval Matrix, we sort each user by item repeat frequencies and choose the Top-M (M=10) users. We tune hyper-parameters using the validation set and terminate training if validation performance doesn’t improve for 20 epochs. Following the common strategy \cite{SASREC}, we pair the ground-truth item with 3 randomly sampled negative items that the user has not interacted with in the validation set and 100 in the test set.  
\subsection{Overall Performance Comparison (RQ1)}
\subsubsection{Comparsion with Baselines}
We report the detailed performance comparison with 10 competitive baselines on different datasets in Table \ref{tbl: baseline} and summarize the observations as follows:
\begin{itemize}[leftmargin=*]
    \item \textbf{TSRec has been empirically validated to achieve state-of-the-art performance. } The results on three datasets indicate that our TSRec method consistently outperforms all types of baselines by a significant margin. Notably, our methodology leads to an improvement of up to $4.03\%$ in HR@5 for the RetailRocket dataset and up to $5.47\%$ in HR@5 for the Diginetica dataset. Our superior performance can be attributed to the following factors: 1) through SRAM, TSRec is able to capture the \textit{sequential pattern} from coarse-grained to fine-grained granularities; 2) with the target-attention-based User-specific Temporal Representation Module, we endow TSRec with the capability of capturing long-range item repeat correlations over time. 3) The Repeat Time Interval Matrix can have multi-perspective time-interval information and the corresponding UTRM is effective to extract the above information.
    \item \textbf{Our method demonstrates robustness across various recommendation scenarios. } Our methodology exhibits significant improvement despite the diverse dataset characteristics, including differences in average sequence length and repeat ratio. Also, the items and users in these three datasets are totally different and this difference will lead to preference characteristics (different distributions of repeat time intervals and different lengths of repeat sequences). This improvement compared with SOTA benchmarks indicates that our method possesses strong generalization capabilities and is robust in practical scenarios.
    \item \textbf{Existing repeat-aware recommendation methods exhibit inferior performance compared to general sequential recommendation techniques.} In the RetailRocket and Diginetica datasets, the evaluated metrics of repeat-aware recommendation methods (e.g., RepeatNet, ReCANet) are significantly lower than those of sequential recommendation methods (e.g., SASRec, BERT4Rec). This outcome may be attributed to the relatively lower repeat ratio (35.03\% and 16.53\%) in these datasets, which causes repeat-aware recommendation methods to overly prioritize repeat items and neglect the sequential relation for new items. In contrast, in the LastFM dataset where the repeat ratio is 89.29\%, the repeat-aware recommendation method can achieve competitive results. In addition, the incorporation of repeat-explore hybrid prediction parts will reduce the representation ability of user historical preference.
\end{itemize}

\begin{figure}[h]
  \centering
  \includegraphics[width=\linewidth]{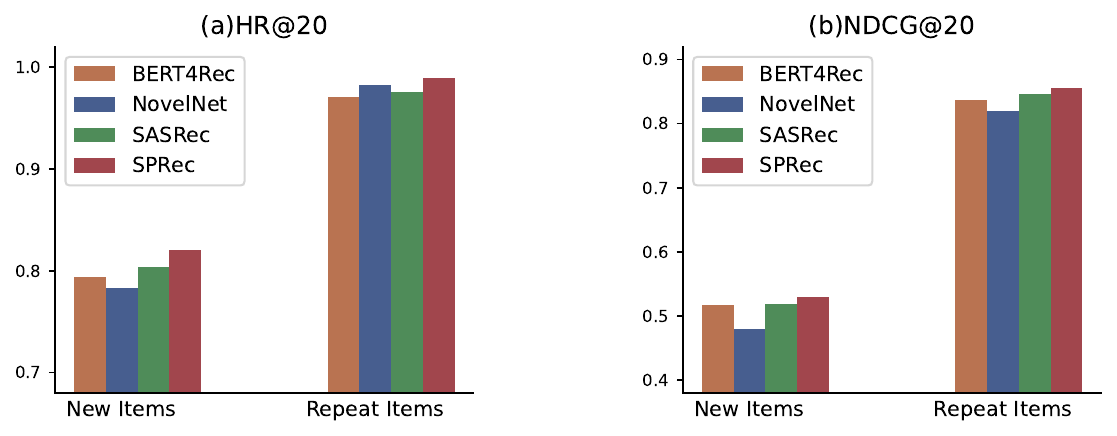}
  \caption{The performance of TSRec is evaluated on both new and repeat items, and compared against various benchmarks on the RetailRocket dataset using HR@20 and NDCG@20 metrics.}
  \label{fig: repeattime}
\end{figure}

\subsubsection{Analysis of Recommending New and Repeat Items} To further compare our method with SOTA benchmarks, we split the prediction accuracy according to the repeat frequency of items - new items and repeat items. In Figure \ref{fig: repeattime}, we present the comparative results with various analogous benchmarks (BERT4Rec, NovelNet, SASRec). Due to spatial constraints, we have only enlisted the commensurate results on the RetailRocket dataset and summarized the following findings:
\begin{itemize}[leftmargin=*]
    \item \textbf{Our proposed method outperforms other benchmarks both in new items and repeat items.} Compared with other recommendation methods, our method performs well in new and repeat items. This finding reveals that our sub-modules can capture fine-grained repeat characteristics of users more effectively, thereby balancing predictions between exploitation and exploration.
    \item \textbf{The approach based on repeat consumption has high accuracy for recommending repeat items, but it leads to a decline in the accuracy rate of new items.} All models perform better for repeat items than for new items because these methods underrate new items as they do not exist in the user's historical sequence. The repeat-aware recommendation methods cannot be compared with sequential recommendation methods for predicting new items. For example, for users who have a preference for purchasing new items, the repeat-aware recommendation methods can not fully predict these users' demands. To summarize, these methods do not strike a balance between exploitation and exploration. 
\end{itemize}

\subsection{Ablation Study(RQ2)}
To validate the effectiveness of different components of TSRec, we remove three components (UTRM, ITRM, and SMM) of TSRec and make a comparison with our TSRec method as shown in Table \ref{tbl:abla}:
\begin{itemize}[leftmargin=*]
    \item \textbf{All the components in the TSRec method are useful to improve the recommendation performance. } Three of these components yield significant drops in performance, indicating their essentiality. More specifically, removing SMM results in an incomplete exploration of the \textit{sequential pattern}, causing the model to only capture basic sequential data, leading to a decline in performance. Additionally, the exclusion of ITRM and UTRM results in a failure to leverage the \textit{temporal pattern}, which hinders the model's ability to predict temporal intentions of repeat consumption.
    \item \textbf{ITRM achieves better performance than UTRM. } Removing ITRM and UTRM will decrease the performance of TSRec, but ITRM is more significant actually. The reason for the performance gap is data sparsity issues in one user repeat sequence. The Repeat Time Interval Matrix has rich \textit{temporal pattern} and the \textit{temporal pattern} from other users is coherent with that of the current user.
\end{itemize}
\begin{table}[]
 \caption{Effectiveness validation of TSRec and its components on three datasets. SMM means Sequence Matching Module; UTRM means User-specific Temporal Representation Module; ITRM means Item-specific Temporal Representation Module. The Best performance is in bold font.}
    \centering
    \renewcommand{\arraystretch}{1.15}
    \resizebox{\linewidth}{!}{
    \begin{tabular}{l|l|ccc|c}
    % \toprule
    \hline  \textbf{ Datasets } & \textbf{ Metric } & \textbf{ w/o SMM} & \textbf{ w/o UTRM} & \textbf{ w/o ITRM} & \textbf{ TSRec}  \\
    \hline  \multirow{2}{*}{\textbf{RetailRocket }}&{ \textbf{HR$@$20 }} & {0.9066} & {0.9061} & {0.9050}& {\textbf{0.9087}}   \\
     {} &{ \textbf{NDCG$@$20 }} & 0.7371 & 0.7365 & 0.7333 & \textbf{0.7523} \\
    \hline  \multirow{2}{*}{ \textbf{LastFM}} & { \textbf{HR$@$20 }} & 0.9466 & {0.9455} & 0.9443 & \textbf{0.9508}  \\
     {} &\textbf{ NDCG$@$20 } & 0.8138 & {0.7913} & 0.8041 & \textbf{0.8152}  \\
     \hline  \multirow{2}{*}{\textbf{ Diginetica}} & \textbf{ HR$@$20 } & {0.9641}& {0.9617} & {0.9598} & \textbf{0.9699}  \\
     {} &\textbf{ NDCG$@$20 } & {0.8127}& {0.8036} & {0.7909} & \textbf{0.8147}  \\
    \bottomrule
    \end{tabular}
    }

    \label{tbl:abla}
\end{table}
\subsection{The Impact of Hyper-parameters(RQ3)}
In the subsequent study, we will examine the effects of the hyper-parameters, namely $L$ and $d$, individually, while preserving the optimal configurations of the remaining hyper-parameters.

\subsubsection{Impact of sequence length $L$.} We validate the robustness of our method and choose different sequence lengths $L$ as shown in Figure \ref{fig:seqlen}. Across the three datasets, it is observed that the truncation of distinct sequences has minimal impact on the HR of the TSRec's prediction,  whereas it has a more notable effect on the final ranking position. This finding indicates that TSRec is robust in accurately identifying relevant items with different sequence lengths. Notably, employing larger sequence lengths may reduce model efficiency. Therefore, to optimize both performance and efficiency, we designate $L=10$ as the hyperparameter for TSRec."
\subsubsection{Impact of embedding dimension $d$.} 
\begin{figure}[htbp]
     \centering
     \begin{subfigure}[t]{0.5\textwidth}
         \centering
         \includegraphics[scale=0.39]{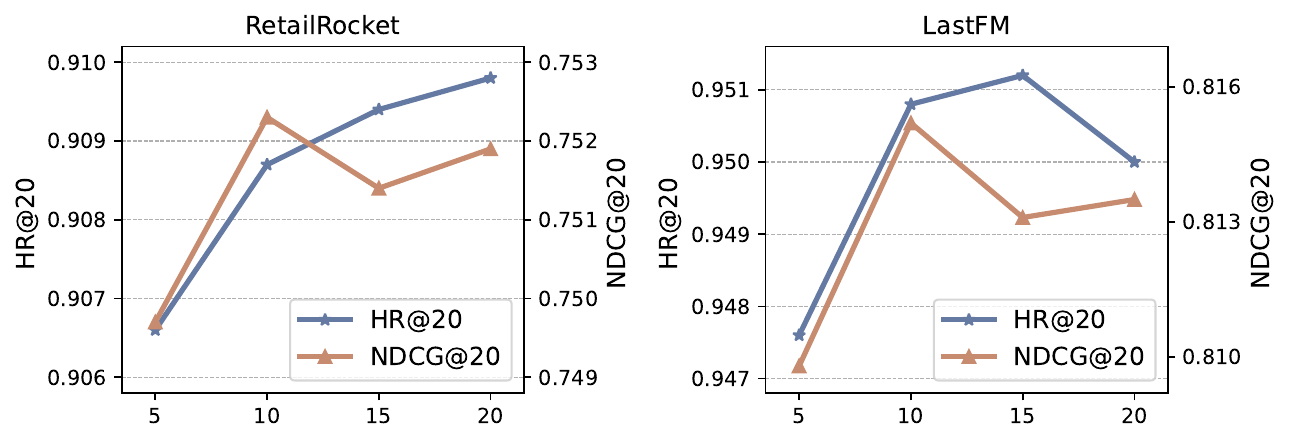}
         \caption{Sequence Length $L$}
         \label{fig:seqlen}
     \end{subfigure}
    \hspace{0.01\textwidth}
     \begin{subfigure}[t]{0.5\textwidth}
         \centering
         \includegraphics[scale=0.39]{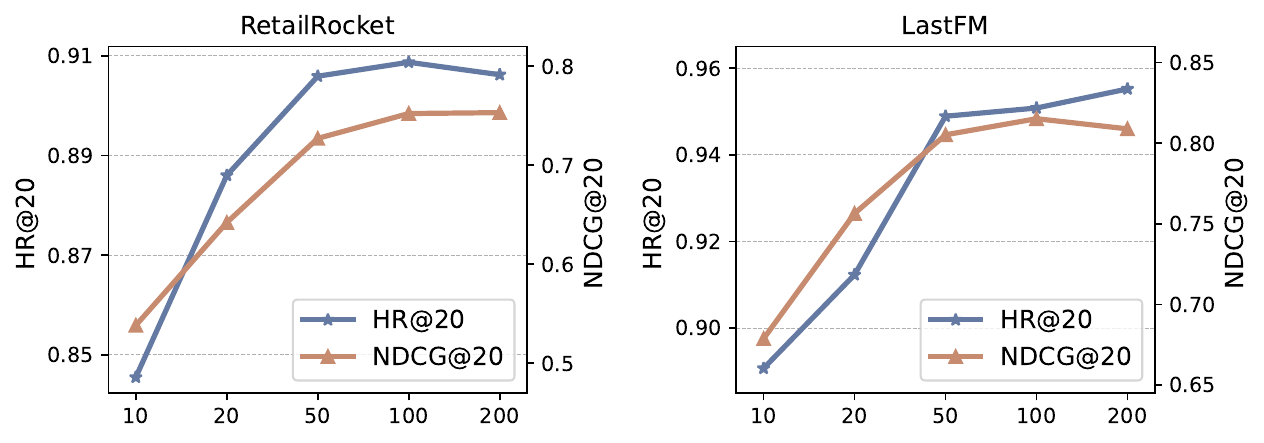}
         \caption{Dimensionality of Vectors $d$}
         \label{fig:embeddim}
     \end{subfigure}
        \caption{The impart of hyper-parameters, namely $L$ and $d$ on two datasets in terms of HR@20 and NDCG@20.}
        \label{fig:abtest}
\end{figure}

The performance of TSRec over a range of embedding dimensions (from 10 to 200) is presented in Figure \ref{fig:embeddim}. Examining the results, we observe that the HR and NDCG of our model initially increase as the embedding dimension expands. Specifically, HR reaches its maximum at the embedding size of 100, owing to the ability of a larger dimension to retain and transmit more information. However, further increasing the embedding dimension will likely lead to overfitting, diminishing the performance of TSRec. As our model achieves satisfactory performance with embedding dimensions of $d \leq 100$, we choose $d = 100$ as the optimal hyper-parameter of TSRec.

\subsection{Case Study of Temporal Modeling(RQ4)}
\begin{figure}[h]
  \centering
  \includegraphics[width=\linewidth]{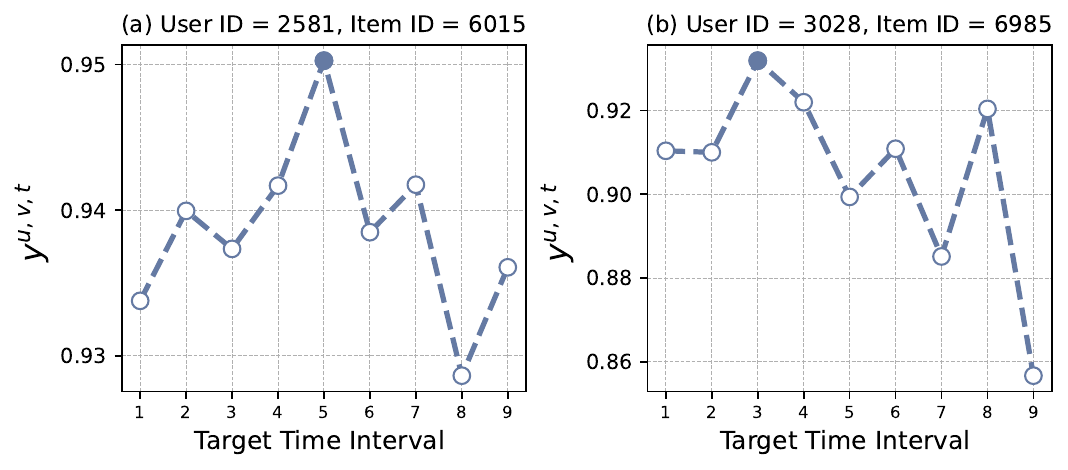}
  \caption{Test cases to verify that TSRec has indeed learned the temporal information of user repeat consumption. Hollow dots represent the target repeat time interval that we virtually generated, while solid dots represent the user's last repeat consumption time interval.}
  \label{fig: case}
\end{figure}

In order to verify that TSRec has learned the temporal information of user repeat consumption, we randomly selected a few test cases. In Figure \ref{fig: case}, the solid dots are the time interval of the user's last repeat consumption while the hollow dots are virtually generated target time intervals around the user's last repeat time interval. Among these dots, we can observe that the solid dots reach the maximum $\boldsymbol{y}^{(u,v,t)}$, and hollow dots have a relatively lower $\boldsymbol{y}^{(u,v,t)}$. This observation indicates that TSRec has indeed learned the relationship between  the time intervals of the user's historical repeat consumption and the target repeat time interval.

\section{Related Work}
\textbf{Sequential Recommendation.} Sequential Recommendation is a task used to predict users' future preferences based on their personalized sequential behaviors. There are various works designed to exploit and model users' history to capture sequential relations and forecast their future behaviors. In these works, Factorized Personalized Markov Chain (FPMC) \cite{FPMC} integrated Matrix Factorization with Markov Chains for next-basket recommendations, embedding the information of adjacent behaviors into item latent factors. \cite{mc1} followed this line and extend it to high-order Markov chains.

We summarized several works most related to our method. To model sequential behaviors, He et al. \cite{fossil} proposed a sparse sequential recommendation approach with Markov chains. Meanwhile, Hidasi et al. proposed GRU4REC \cite{GRU4REC, GRU4REC+} utilizing RNNs and an improved pairwise loss to model short-term preferences for session-based recommendations. Additionally, Li et al. proposed NARM \cite{NARM} leveraging a global and local encoder with an attention module to model both user short- and long-term interest. Furthermore, other works employed memory networks \cite{MANN} and CNNs \cite{CASER} for sequential recommendations. Meanwhile, SASRec \cite{SASREC} and BERT4Rec\cite{BERT4REC} followed the structures of Transformer \cite{Transformer} and BERT \cite{bert} to bridge the item-item relationship and gain the benefits in the next-item prediction task. As these works are similar to our repeat-aware recommendation approach, we have selected several crucial or state-of-the-art recommendation models as comparative baselines.

\textbf{Repeat-aware Recommendation.} The repeat consumption has been extensively investigated in areas such as music listening \cite{ACT-R, musicrec}, E-commerce \cite{SLRC}, grocery shopping \cite{ReCANet}, online novel reading \cite{NovelNet}, and live streaming \cite{livestream}. In these fields, repeat consumption has been a vital indicator for modeling repeated user-item interaction.
In recent years, there has been a series of studies that focus on modeling intrinsic patterns related to repeat consumption through simplified tasks. For instance, Anderson et al. \cite{recency} analyzed repeat consumption patterns and behaviors and identified recency as a crucial factor in predicting user purchasing patterns through a cache-based method. Their findings suggest that users tend to purchase recently-purchased items. Meanwhile, Benson et al. \cite{bore} demonstrated the impact of users getting bored with an item by analyzing the temporal gaps between users' consecutive purchases of identical products. They found that greater temporal gaps between purchases indicated user boredom with the item, which eventually leads to abandonment. Wang et al. proposed SLRC \cite{SLRC} exploring the short-term effect and the lifetime effect to model repeat consumption by integrating Collaborative Filtering and Hawkes Process methods. While these approaches mainly model repeat consumption by exploring its characteristic, some studies have considered the design of model structures. For example, Ren et al. proposed RepeatNet \cite{RepeatNet}, which used the copy attention mechanism to predict the probability of repetition and exploration. Ariannezhad et al. proposed ReCANet \cite{ReCANet}, a framework with LSTM layers to explicitly model the repeat behavior of users. Li et al. proposed NovelNet \cite{NovelNet} which used a pointer network combined with a pointwise loss to model user repeat consumption behavior. Though the above works have considered modeling repeat consumption, they overlooked the fine-grained characteristics behind repeat consumption, which might lead to the degradation of model prediction performance.

\section{Conclusion}
In this paper, we investigate various inherent characteristics to enhance the repeat-aware recommendation. A novel approach called \textbf{T}emporal and \textbf{S}equential repeat-aware \textbf{Rec}ommendation (\textbf{TSRec} for short) is presented to capture the \textit{temporal pattern} and \textit{sequential pattern}. Offline experimental results on three public datasets demonstrate the superiority of TSRec over several state-of-the-art methods for both new and repeat items. Ablation studies illustrate the effectiveness of the three components of TSRec. Additionally, case studies demonstrate TSRec's ability to capture meaningful temporal information via time interval modeling. The phenomenon of repeat consumption is prevalent in many recommendation scenarios (e.g., e-commerce, music, and social media). The temporal and sequential patterns underlying repeat consumption represent the inherent characteristics of user repeat purchase behavior. We believe that modeling these patterns holds significant potential to enhance the performance of the recommendation system.

\begin{acks}
To Robert, for the bagels and explaining CMYK and color spaces.
\end{acks}

%%
%% The next two lines define the bibliography style to be used, and
%% the bibliography file.
% \newpage
\bibliographystyle{ACM-Reference-Format}
\balance
\bibliography{sample-base}

%%
%% If your work has an appendix, this is the place to put it.
% \appendix

% \section{Research Methods}

% \subsection{Part One}

\end{document}